\documentclass[twocolumn,showpacs,preprintnumbers,amsmath,amssymb]{revtex4}

\usepackage{graphicx}

\begin{document}

\title{Switching a spin-valve back and forth by current-induced domain wall motion}

\author{J. Grollier, P. Boulenc, V. Cros}
\affiliation{Unit\'e Mixte de Physique CNRS/THALES, Domaine de
Corbeville, 91404 Orsay, and Universit\'e Paris-Sud, 91405 Orsay,
France}
\author{A. Hamzi\'{c}}
 \altaffiliation[~on leave from ] {the Department of Physics,
Faculty of Science, HR-10002 Zagreb, Croatia.}
 \affiliation{Unit\'e Mixte de Physique CNRS/THALES, Domaine de
Corbeville, 91404 Orsay, and Universit\'e Paris-Sud, 91405 Orsay,
France}
\author{A. Vaur\`es}
\affiliation{Unit\'e Mixte de Physique CNRS/THALES, Domaine de
Corbeville, 91404 Orsay, and Universit\'e Paris-Sud, 91405 Orsay,
France}
\author{A. Fert}
\email{albert.fert@thalesgroup.com} \affiliation{Unit\'e Mixte de
Physique CNRS/THALES, Domaine de Corbeville, 91404 Orsay, and
Universit\'e Paris-Sud, 91405 Orsay, France}

\author{G. Faini}
\affiliation{Laboratoire de Photonique et de Nanostructures,
LPN-CNRS, Route de Nozay, 91460 Marcoussis, France}

\begin{abstract}
We have studied the current-induced displacement of a domain wall
(DW) in the permalloy (Py) layer of a Co/Cu/Py spin valve
structure at zero and very small applied field. The displacement
is in opposite direction for opposite dc currents, and the current
density required to move DW is only of the order of
10$^6$~A/cm$^2$. For $H$ = 3~Oe, a back and forth DW motion
between two stable positions is observed. We also discuss the
effect of an applied field on the DW motion.
\end{abstract}

\maketitle

Switching the magnetic configuration of a micro-device by spin
transfer from a spin-polarized current, rather than by applying an
external field, is the central idea of a present extensive
research~\cite{Slonczewski, Berger1, Katine, Sun, Grollier1,
Wegrowe, Tsoi}.~In 1996, Slonczewski~\cite{Slonczewski} showed
that the magnetic moment of a ferromagnetic layer can be reversed
by injecting a spin-polarized current into this layer.~This
prediction has been convincingly confirmed by series of
experiments on pillar-shaped magnetic multilayers~\cite{Katine,
Sun, Grollier1}, nanowires~\cite{Wegrowe} or
nano-contacts~\cite{Tsoi}. However, the current density required
in the existing experiments is relatively high, of the order of
10$^7$~A/cm$^2$, and some reduction of this density is necessary
for practical applications.

Another way to change a magnetic configuration is by
current-induced motion of a domain wall (DW). DW-drag by a current
has been predicted by Berger~\cite{Berger2} and its theory has
been recently revisited by Waintal and Viret~\cite{Waintal}.~When
a spin-polarized current goes through a DW, the torque, resulting
from the interaction of the conduction electron spins with the
exchange field in the DW, progressively rotates the spin
polarization of the current. Reciprocally, the spin-polarized
current exerts an exchange torque on the magnetization within the
DW, which is the origin of the DW motion predicted by
Berger~\cite{Berger2}.

Freitas and Berger~\cite{Freitas} have obtained some experimental
evidence of DW-drag by using Kerr microscopy to detect the DW
position. In recent similar experiments  Gan \textit{et
al.}~\cite{Gan} have also measured DW displacement due to current
pulses by imaging the DW by MFM before and after current
pulses.~The main features in these two sets of experiments are
that the direction of the the DW motion is reversed when the
direction of the current pulses is reversed, and that the order of
magnitude of the current pulses needed to move the DW is about
10$^7$~A/cm$^2$.

In recent experiments on Co/Cu/Py spin valves~\cite{Grollier2}, we
have also found that a dc current can switch the magnetic
configuration of the spin valve by moving a DW in the permalloy
(Py) free layer. We observed that a DW can be moved away from an
artificial  pinning center (notch) when the current density
exceeds a threshold value of the order of
10$^7$~A/cm$^2$.~However, the mechanism of the DW displacement was
not completely clear.~In fact, the DW could not be displaced at
zero field, but only by combining current and applied field.~Also,
the motion direction was determined by the field direction and not
reversed when the current was reversed.~These results suggested a
more complex mechanism than described by Berger~\cite{Berger2},
with a possible effect of the applied field on the DW distorted by
the current.~In this paper we present much clearer results
obtained on spin valves but with weaker DW pinning.~The
displacement of the DW is obtained at zero field, in opposite
directions for opposite current directions and with definitely
lower current densities.

Our samples are 300~nm wide and 20~$\mu$m long stripes patterned
by e-beam lithography using a lift-off technique.~The spin valves,
deposited by sputtering, have a final structure
~CoO(30~$\AA$)/Co(70~$\AA$)/Cu(100~$\AA$)/Py(50~$\AA$)/Au(30~$\AA$).
The top Au electrodes are processed by UV lithography. In contrast
with our previous experiments~\cite{Grollier2}, the only pinning
centers for the DW in the Py soft layer are natural defects of the
stripe. All the measurements were performed at room temperature.

\begin{figure}
  \includegraphics*[width=8 cm]{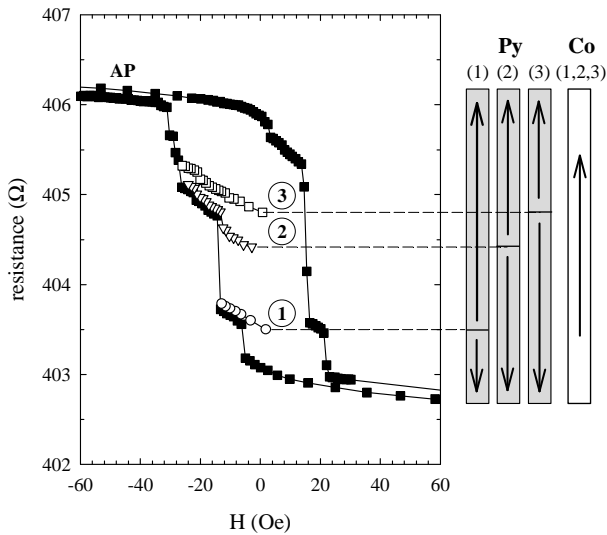}\\
  \caption{($\blacksquare$)~-~GMR minor cycle associated with the reversal of the permalloy layer of the Co/Cu/Py trilayer at $T$ = 300~K. The field is
  applied along the stripe. The magnetization of the Co layer is pinned in the positive field direction.
($\square$,~$\triangledown$,~$\circ$)~-~variation of the
resistance when the cycle is stopped at one of the plateaus and
the field is brought back to zero.~Also sketched are the DW
position in
  the Py stripe and the magnetic configuration corresponding to the
  levels 1, 2 and 3).} \label{fig:fig1}
\end{figure}

In Fig.~\ref{fig:fig1} we show a typical GMR minor cycle
associated with the reversal of the magnetization in the Py layer,
i.e.~with the motion of a DW   from one end of the Py stripe to
the other one. The plateaus are due to the pinning of the DW on
defects in the Py stripe. We emphasize that the series of plateaus
on the GMR curve is highly reproducible. As shown for the left
half of the cycle, a DW remains pinned on the same defect when the
field is brought back to zero. We can therefore start an
experiment at zero field with the DW pinned in one of the three
positions (sketched) corresponding to the resistance levels 1, 2
and 3.~The results presented below correspond to experiments
performed with a DW initially pinned in the configuration 2.~The
first series of experiments are performed by varying the current
at zero or very low field (parallel to the stripe). In another
series of experiments, we study the influence of a larger bias
field on the current-induced DW motion.

\begin{figure}
  \includegraphics*[width=7 cm]{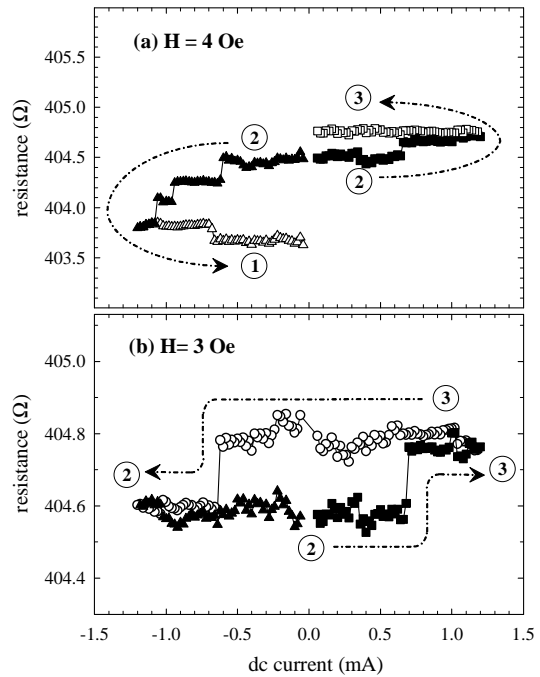}\\
  \caption{Resistance \textit{vs.} current in very low constant field $H$ along
  the stripe.
   (a) $H$ = 4~Oe ($\blacksquare$~-~motion from 2 to 3 with a positive current; $\blacktriangle$~-~motion from 2 to 1 with a negative current); (b)
$H$ = 3~Oe (motion from 2 to 3 with a positive current and back to
2 with a negative current).~The numbers 1, 2 and 3 refer to the DW
configurations and corresponding resistance levels of
Fig.~\ref{fig:fig1}. A small contribution ($\sim I^2$), due to the
Joule heating ($\triangle T\simeq$ 5~K), has been subtracted for
clarity.} \label{fig:fig2}
\end{figure}

In Fig.~\ref{fig:fig2} we present results obtained by varying the
dc current at constant field close to zero (4 and 3 Oe). As shown
in Fig.~\ref{fig:fig2}a, starting from the DW in position 2, we
can move the DW to position 3 by increasing the current above the
positive critical value $j^{+}_{c2}$(4~Oe)~=~+~0.65~mA and
decreasing it back to zero.~Alternatively, the DW is moved in the
opposite direction (from 2 to 1) with a negative current exceeding
$j^{-}_{c2}$(4~Oe)~=~$-$ 1.1~mA (in our notation $j^{+}_{cn}$ and
$j^{-}_{cn}$ are the critical currents required to move the DW
from position \textit{n} to positions \textit{n+1} and
\textit{n-1} respectively). The same type of behavior is observed
for all applied fields between 0 and 7~Oe. However, even in this
very small field range, there is some field dependence of the
critical currents: $j^{+}_{cn}$($H$) ($j^{-}_{cn}$($H$)) decrease
when $H$ decreases (increases) and favors a DW motion from
\textit{n} to \textit{n+1}(\textit{n-1}).

Fig.~\ref{fig:fig2}b presents an example of back and forth DW
motion, namely the motion from 2 to 3 with positive dc current and
a return to 2 with a negative dc current. The obvious conditions
for this back and forth motion are $j^{+}_{c2}(H) < j^{+}_{c3}(H)$
(required to stop the first motion in configuration 3) and $\left|
j^{-}_{c2}(H) \right|
> \left| j^{-}_{c3}(H) \right|$ (necessary for the return to the
configuration 2). It turns out that these conditions are satisfied
for the pinning centers 2 and 3 of our sample, at least for $H$ =
3~Oe.

The behavior observed in the field range close to zero
(approximately, 0 $< H <$ 7~Oe) can be summarized as follows. A DW
can be displaced between pinning centers and, in agreement with
what is predicted for a displacement by Berger's
mechanism~\cite{Berger2}, its motion is in opposite directions for
opposite currents. The dc current density needed to  move the DW
is of the order of 10$^6$~A/cm$^2$, that is an order of magnitude
smaller than the currents required for the magnetization reversal
in pillar-shaped multilayers~\cite{Katine, Sun, Grollier1}. There
is however some uncertainty in the exact value of the current
density in Py. If the electron mean free paths in Py and Co were
much smaller than the thicknesses of the Py and Co layers (which
is far from being satisfied at room temperature) and if we could
also neglect the specular reflections of the electrons at the
interfaces, there would be an uniform current density in the
multilayer~\cite{Barthelemy}, that is, for example,
8$\cdot$10$^6$~A/cm$^2$ for 0.6~mA. In the opposite limit of
almost independent conduction by the magnetic and nonmagnetic
layers (this would correspond to layer thicknesses larger than the
mean free paths, or also to almost complete specular reflections
at the interfaces, with, in both cases, a vanishing GMR), a
straightforward calculation, based on the resistivity of the
different metals at room temperature, leads to a current density
of 1.75~$\cdot$10$^6$~A/cm$^2$ in Py. In an intermediate situation
(mean free path of the order of the Co and Py thickness,
consistently with the small but non-zero GMR, and certainly some
current channeling in Cu by specular reflections), the real
current density in Py is probably in-between, that is of the order
of few 10$^6$~A/cm$^2$.

\begin{figure}
  \includegraphics*[width=7 cm]{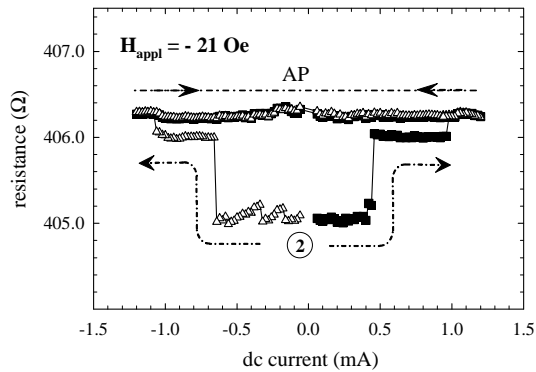}\\
  \caption{Resistance {\it vs.}  current  for $H$ = $-$~21~Oe.} \label{fig:fig3}
\end{figure}

Out of the low field range described above, the behavior becomes
more complex. An example of experimental result is shown in
Fig.~\ref{fig:fig3} for $H$ = $-$~21~Oe favoring an antiparallel
(AP) configuration.~A positive current moves the DW from position
2 to the end of the stripe (AP resistance level), which is
consistent with the motion direction induced by a positive current
at low field. On the other hand, in contrast with the low field
behavior, the motion is not reversed for negative currents and the
final state is still the AP configuration. For positive fields out
of the low field range, the same type of behavior is observed,
with a motion towards a more parallel configuration. We can
therefore conclude that, out of the low field range, the current
is still able to unpin the DW, but the direction of the DW motion
is now controlled by the applied field direction.~This behavior
was also observed in our former experiments~\cite{Grollier2} with
stronger (artificial) pinning centers, where the motion could be
obtained only by combining current and applied field.

We will now focus on the interpretation of the DW-drag effects at
zero or low field.~To start with, we can rule out any contribution
from Joule heating.~From the small quadratic resistance increase
with current (the term subtracted in Figs.~\ref{fig:fig2}
and~\ref{fig:fig3}), the maximum increase of $T$ is about 5 K, and
we have checked that, at 300~K, this has practically no effect on
the GMR minor loop.~An even stronger argument is that heating
could not explain that opposite currents produce motions in
opposite directions. The Oersted field generated by the current
($\leq$~20~Oe), in a perfect structure, is in the plane of the DW,
and it cannot favor a motion in one or in the other direction. In
the presence of defects, the Oersted field might have a component
out of the DW plane, but it can be hardly imagined that different
defects give always the same direction for this component and the
DW motion.~Only the spin transfer mechanism, first proposed by
Berger, is consistent with the experimental results at zero or
very small field and, particularly, can explain the reversal of
the motion with opposite currents. Berger~\cite{Berger2} expresses
the spin transfer by a torque corresponding to the field
$H_{B}=jP\hbar/e\delta M_{s}$, perpendicular to the layer. With
$M_{\mathrm{s}}$ = 860~$\cdot$~10$^3$ A/m, $\delta$ (DW thickness)
= 100 nm, P (polarization) = 1 and $j$ = 5~$\cdot$~10$^6$
A/cm$^2$, we obtain $H_{\mathrm{B}} \simeq$ 3.8~Oe, just in the
range of the pinning fields of the DW in the Py layer.~Although,
at the present stage of the theory, the connection between the
torque of $H_B$ and the value of the critical current is still
unclear, we can conclude that the DW motions at zero or very low
field, characterized by a reversal of the motion in opposite
currents, can be ascribed to a spin transfer mechanism. On the
other hand, the behavior observed at higher fields, with combined
influence of current and applied field, is more complex.~As
suggested by the model of Waintal and Viret~\cite{Waintal}, it
could be that, in this regime, the depinning of the DW is induced
by the distortion of the wall, while the succeeding motion is
predominately driven by the field.

In conclusion, we have presented experiments in which a spin valve
is  switched by current-induced DW motion. In zero or very low
field, the DW displacement is in opposite directions for opposite
dc currents, and back and forth motions between two pinning
centers can be obtained. Our results are consistent with the spin
transfer mechanism introduced by Berger~\cite{Berger2}.~A more
complex and unclear behavior is observed when the effect of the
current is combined with the effect of an applied field.

This work was supported by the EU through the RTN "Computational
Magnoelectronics" (HPRN-CT-2000-00143) and the Minist\`ere de la
Recherche et de la Technologie through the MRT "Magmem II"
(01V0030) and the ACI contract "BASIC" (27-01).

\end{document}